\documentstyle[prl,multicol,aps]{revtex}
\begin{document} 
\draft
\twocolumn[\hsize\textwidth\columnwidth\hsize\csname@twocolumnfalse\endcsname

\title{ Statistical Study for Eigenfunctions of  
1-dimensional Tight Binding Model
}

\author{Wen-ge Wang$^{[a,b]}$ and Bambi Hu$^{[a,c]}$}
\address{ 
$^a$ Department of Physics and Centre for Nonlinear Studies, 
Hong Kong Baptist University, Hong Kong, China 
\\ $^b$ Department of Physics, South-east University, Nanjing 
210096, China 
\\ $^c$ Department of Physics, University of Houston, TX77204, USA}
 
\maketitle
\widetext
\begin{abstract} 
\begin{center}
\parbox{14cm}{
For energy eigenfunctions of 
1-dimensional tight binding model, 
the distribution of ratio of 
their nearest components, denoted 
by $f(p)$, gives information for 
their fluctuation properties. 
 The shape of $f(p)$ is studied numerically 
for three versions of the 1D tight binding model. It is 
found that when perturbation is strong the shape of 
$f(p)$ is usually quite close to that of the Lorentzian 
distribution. In the case of weak perturbation 
the shape of the central part of $f(p)$ is model-dependent 
while tails of $f(p)$ are 
still close to the Lorentzian form. 
}
\end{center}

\end{abstract}
]
\narrowtext

\begin{center} \Large I. Introduction \rm \end{center}

    Different versions of the one dimensional tight binding model with 
constant near-neighbor-hopping have 
attracted lots of attention for years (e.g., see 
\cite{Thouless,Sokoloff-rep}).  Although 
these versions of the model has the common feature 
of  Hamiltonian matrices being tridiagonal, 
they have been found   
showing distinct global behaviors such as localization 
and delocalization of energy eigenfunctions. 
For example, 
 for the Anderson model \cite{And} with random 
diagonal elements for the Hamiltonian matrix, 
localization can be established rigorously \cite{Ishii}. 
For the case of diagonal elements being quasi-periodic 
 as in the 1D quasi-periodic tight binding model, both 
localized and extended wavefunctions have been found   
\cite{Sokoloff-rep,Aubry,Soukoulis,Kohmoto92}. 
Recently, another interesting version 
of the model has also been studied. 
It has been found that, if some correlation 
is introduced into the diagonal 
elements of the Hamiltonian matrix  
of the Anderson model, delocalization 
can appear \cite{Flores,Dunlap,Heinrichs,Sanchez,Evangelou}. 
Clearly,  global properties as localization and delocalization 
are not determined merely
by  the tridiagonal structure of the Hamiltonian 
matrices. 

    The main difference between the 
Anderson model and the quasi-periodic 
tight binding model 
 lies in the global properties of the diagonal elements 
 of their Hamiltonian matrices. More exactly, it lies in 
whether the diagonal elements  are  
random or quasi-periodic (quasi-random). 
Since locally there is 
no difference between random and quasi-random numbers, one can 
expect that the difference between local 
statistical properties of energy  
 eigenfunctions for the two models  
 is not so distinct as that between their 
global behaviors.   

    Statistical properties of eigenfunctions have been 
studied extensively in recent years (for averaged 
shape see, e.g., 
\cite{CCGI96,WIC97,FGGK94,CFI97,BGIC97,ZBFH,FI97Nov}, 
for fluctuations and correlations see, e.g., 
\cite{CFI97,Eckhardt,XGWYF95,Prigodin,Hortikar}). 
It is known that the Random Matrix Model (RMM) 
(see, e.g., \cite{Mehta,GMW}) and the Band Random 
Matrix Model (BRMM)
(see, e.g., \cite{FM94,I95})  can be used to describe 
quantum chaotic systems and quasi-1D disordered models,
respectively. Although one can expect that some features of 
the above mentioned 1D tight binding model may be described 
by the BRMM with band width equal to 1, 
due to the  constant near-neighbor-hopping, 
it is impossible for all its important
features to be able to 
 be described by the BRMM (let 
alone RM). Therefore, when studying statistical 
properties of the 1D tight binding model one must 
keep in mind its peculiar features.

    When studying statistical properties of 
energy eigenfunctions 
    the first quantity coming to one's mind
may be the distribution of the components 
$c_i$ of the eigenfunctions. 
However, for the 1D Anderson model, 
due to the exponential decay of $|c_i|$ from its 
maximum, the distribution of $c_i$ 
does not provide much information 
for local properties. On the other hand, the distribution of 
$p_i$, for the so-called Riccati variable 
$p_i \equiv c_i/c_{i-1}$, is related to both local 
 and global properties of eigenfunctions \cite{Luck}. 
In this paper we will 
study properties of this distribution.

    This paper has the following structure. In section II, 
some properties of $p_i$ are discussed. It is shown that 
$f(p)$, 
the distribution of $p_i$, not only gives information for  local 
statistical properties of  energy 
eigenfunctions, but also is related to 
 a global property, the Lyapunov exponent. 
Tails of $f(p)$ are expected to behave as $1/p^2$ and 
in some cases 
the shape of  $f(p)$ is expected to be close 
to the Lorentzian distribution. 
In section III, the shape of $f(p)$ is studied numerically 
for the Anderson model. It is shown that in the case of 
weak disorder most of $f(p)$ can be fitted quite 
well by the Lorentzian distribution. In the case of strong 
disorder, central parts of $f(p)$ deviate notably from the  
Lorentzian form and some features of $f(p)$ can be explained 
by perturbation theory. Section IV is devoted to 
a discussion about the shape of $f(p)$ for the other two versions 
of the 1D tight binding model mentioned above.
Numerically, more or less similar results as for the 
Anderson model have been found for the two models. 
A brief conclusion and discussion are given in section V.

\begin{center}  \Large II. Properties of $p_i$  \rm \end{center}

     The Hamiltonian matrix of the 1D tight binding model
 is of the form 
\begin{equation} H_{ij}= \epsilon _i \delta _{ij} 
+ v (\delta _{i,j+1} + \delta _{i,j-1}). 
\label{Hele}  \end{equation}
Without the loss of generality, we take $i,j=0,1,2, \cdots N$. 
When $N$ goes to infinity one has a chain of infinite length. 
Corresponding to the above  mentioned three versions of the model,  
the site energies $\epsilon_i$  are random numbers, 
quasi-periodic  numbers, and partly
random numbers, respectively,  
distributed  in the region $[-w/2,w/2]$.
For the first and third versions flat 
distributions will be chosen. 
Without the loss of generality, we will take $w=1$
in numerical calculations. 

    The  stationary Schr\"{o}dinger 
equation for eigenenergy $E$  is
\begin{equation} 
vc_{i+1} + vc_{i-1} + \epsilon_i c_i = E c_i  
\label{sch} \end{equation}
with $c_k, k=i-1,i,i+1,$ being components of the 
corresponding eigenfunction. 
Dividing both sides of Eq. (\ref{sch}) by $vc_i$, one 
can see that it is convenient to introduce another 
quantity $p_i \equiv c_i/c_{i-1}$, for which  Eq. 
(\ref{sch}) becomes 
\begin{equation} p_{i+1} = \alpha _i - 
\frac 1{ p_i} \label{pi} 
\end{equation} 
where 
\begin{equation} \alpha _i = \frac{E- \epsilon_i}{v}. 
\label{alpha} \end{equation}
Notice that the value of $p_1$ is determined by 
Eq. (\ref{sch}) for $i=0$ ($c_{-1}$ does not exist), i.e., 
 $p_1 = \alpha _0$. As a result, once the eigenenergy $E$ is given, 
the eigenfunction can be obtained immediately from Eq. (\ref{pi})
for $i=1, \cdots N-1$. Eq. (\ref{pi}) for $i=N$ (without 
$p_{N+1}$ which does not exist), i.e., $p_N=1/ \alpha _N$, 
is the condition for $E$ to be an eigenenergy. Conversely, 
one can also calculate the eigenfunction from $p_N= 
1/ \alpha _N$ and take $p_1=\alpha _0$ as the condition 
for $E$ to be an eigenenergy. In either way, 
 even for an arbitrary value of $E$  
one can calculate a sequence of $p_i$ from Eq. (\ref{pi}) 
(neglecting the last equation). 

    A relation between $p_i$ 
and $p_1$ can be get from Eq. (\ref{pi}), 
\begin{equation} 
p_i = \frac{A_i p_1 + B_i}{A_{i-1} p_1 + B_{i-1}}, 
\label{p1pi} \end{equation}
where $A_i$ and $B_i$ are determined by 
recurrence relations   
\begin{equation} A_i =\alpha _{i-1} A_{i-1} - A_{i-2} 
\label{Arec} \end{equation}
and 
\begin{equation} B_i =\alpha _{i-1} B_{i-1} - B_{i-2},
\label{Brec} \end{equation}
respectively, with 
\begin{equation} A_2=\alpha _1, \ \ A_1 = 1 \end{equation}
and
\begin{equation} B_2=-1, \ \ \ \ B_1=0.  \end{equation}
(For more detailed properties of $A_i$ and $B_i$, see Appendix 
A.)

    For a sequence of $p_i$ obtained from Eq. (\ref{pi})
for  $i$ from 1 to $N$, 
when $N$ goes to infinity, one can 
calculate the distribution 
of $p_i$, denoted by $f(p)$. 
For a finite $N$, only a histogram can be obtained. 
For the top of the histogram to be close 
to  a smooth function, $N$ must be large. 
For  brevity, we also use $f(p)$ to 
indicate the distribution for a finite sequence of $p_i$.  
The distribution $f(p)$ gives information for 
not only local fluctuation properties 
but also some global properties of eigenfunctions. For example, 
the Lyapunov exponent can be expressed as 
\[ \gamma (E) = \lim _{n \to \infty } \frac 1n 
\ln |\frac{c_n}{c_0}|  
= \lim _{n \to \infty } \frac 1n \sum_{i=1}^n \ln |p_i| \] 
\begin{equation} 
=\int f(p) \ln |p| dp.  
\label{Lyap} \end{equation} 

    As is known, for a localized eigenfunction, $|c_i|$ 
usually first increases (on average) 
exponentially and then decreases (on average) 
exponentially as $i$ varies from $0$ to $N$. Let us 
define the Lyapunov exponent for a finite sequence of 
$p_i$ with $i$ from $j+1$ to $k$ as 
\begin{equation} 
 \gamma (E,\beta ) =  \frac 1{k-j} 
\ln |\frac{c_k}{c_{j}}|  
=  \frac 1{k-j} \sum_{i=j+1}^k \ln |p_i| 
\label{Lyap2} \end{equation}
where $\beta $ denotes the sequence of $p_i$. 
Denoting the sequence of $p_i$ of the increasing part 
of the eigenfunction as  $\beta_+$ and that of the decreasing 
part as  $\beta_-$, one can see from Eq. (\ref{Lyap2})    
that $\gamma(E,\beta_+) > 0$ 
and $\gamma(E,\beta_-) < 0$. Clearly, $f_+(p)$, the distribution 
of $p_i$ for $\beta _+$, and $f_-(p)$, the distribution 
of $p_i$ for  $\beta_-$, can not be equal to each other 
even when $N$ goes to infinity. 
Therefore, when studying $f(p)$ distributions, there 
are three cases that should be distinguished, 
i.e., $f_+(p)$, $f_-(p)$ and $f_a(p)=(f_+(p) + 
f_-(p))$ for $p_i$ in both $\beta _+$ and 
$\beta _-$. (In what follows we use $f(p)$ 
to denote all the three cases.) 

    Since it is difficult to study the shape of $f(p)$ 
analytically and directly, a related problem can be 
studied to gain some reasonable 
predictions for behaviors of $f(p)$. 
Suppose one has a set of $p_1$ with a distribution 
$F_1(p_1)$. Making use of Eq. (\ref{pi}), one can change 
$p_1$ to $p_2$ and  
 obtain the distribution for $p_2$, 
denoted by $F_2(p_2)$. Proceeding this procedure, one can 
obtain the distribution for $p_n$, $F_n(p_n)$. As will 
be shown in Appendix B, $F_n(p_n)$ does not approach any 
fixed distribution when $n$ goes to infinity, but 
their tails behave as $1/p^2$ and tails of $f(p)$ should 
also decay as $1/p^2$. 

    What is also worth mentioning here is that the Lorentzian 
distribution 
\begin{equation} f_L(p)= \frac{a/\pi}{(p-b)^2+ a^2}
\label{lorentz} \end{equation} 
with
\[ b=\frac {\alpha _i}2, \ \ \ \ \ a=\sqrt{1-b^2} \] 
is invariant under the change of the variable from $p$ 
to $(\alpha _i -1/p)$ when $|\alpha _i| <2$. Since 
\begin{equation} \int ^{\infty }_{-\infty } f_L(p)
\ln |p| dp =0 \end{equation} 
(see Appendix B), it is impossible for 
$f_{\pm }(p)$ to be 
equal to $f_L(p)$ for localized states,
while it is possible for $f_a(p)$. However, when $\gamma 
(E)$ is small, it is possible for $f_{\pm }(p)$ 
to be close to $f_L(p)$.

\begin{center} {\Large III. $f(p)$ for the Anderson model}
\end{center}

    In this section we study the form of $f(p)$ 
for the Anderson model by numerical 
calculations. 
First, in subsection A we discuss the 
methods for calculating $f(p)$. Then, in 
subsection B we study the shape of $f(p)$ for 
 the case of weak disorder, 
i.e., when perturbation is strong. 
Finally, in subsection C 
the form of $f(p)$ is discussed for the case of strong 
disorder. 

\begin{center} {\bf A. Methods for calculating $f(p)$}
\end{center} 

  There are three methods for calculating $f(p)$ 
numerically. For the first method, one can try to find 
out eigenfunctions of a Hamiltonian matrix 
with $N$ large enough for a good statistics 
for $f(p)$ obtained from the data of
one energy eigenfunction. 
Since when diagonalizing large matrices 
the so-called  $QR$ method usually 
does not give correct results for long tails of 
eigenfunctions with very small $|c_i|$,  we have used 
the following method to calculate eigenfunctions. 
First we chose an initial value $E'$ and calculated 
$p_1, p_2, \cdots p_{N}$ by making use of  
Eq. (\ref{pi}) for $i=0,1,\cdots N-1$.  
 Then, we changed $E'$ to make $p_N$ 
satisfy Eq. (\ref{pi}) for $i=N$ 
as exactly as possible, i.e., 
to reduce the value of $err=|(p_N-1/\alpha _N)/p_N|$
as much as possible. 
As an example, when 
 $v=4.0$ and $N=5 \times 10^{5}$,  such an 
eigenfunction has been found 
with eigenenergy $E \approx -4.0044$.
The value of $err$ for this eigenfunction 
is about $10^{-14}$. 
For this eigenfunction, $|c_i|$, on average, increases 
exponentially from $i=0$ to $i=N$. 
The $f_+(p)$ distribution for this eigenfunction 
is given in  Fig.1a (circles).
   The line in Fig.1a gives $f_-(p)$ for 
another eigenfunction 
with $E \approx -4.002$, for which $|c_i|$ decreases, on 
average, exponentially from $i=0$ to $i=N$. 
This eigenfunction was obtained by calculating the 
sequence of $p_N, p_{N-1}, \cdots p_1$ and taking 
Eq. (\ref{pi}) for $i=0$ as the condition for 
eigensolution. The value of
$eer$ for this eigenfunction is also about $10^{-14}$. 
The shortcoming of this method is that it 
is difficult (if not impossible) to find 
out eigenfunctions for which both the increasing 
and decreasing (on average) parts of 
$|c_i|$ are large. 

  For the second method for calculating 
$f(p)$, one can 
diagonalize (many) Hamiltonian matrices 
with the same value of $v$ but  
 different realizations 
of the random diagonal elements. There are two restrictions 
for the dimension $N$ of the matrices. One 
is that it must be large enough for eigenfunctions 
to be localized (exponentially decay). 
The other is that the QR method for diagonalization 
gives correct results for long tails of 
eigenfunctions.   
For eigenfunctions thus obtained with 
eigenenergies in a small energy region
one can calculate a $f_a(p)$ 
distribution. The shortcoming of this method 
is that when $v$ is large, in order to 
obtain localized eigenfunctions the dimension 
$N$ of the matrices must also be large, which  
requires powful computers and long calculation 
time. 

  The third method is suggested by the fact 
that to obtain a sequence of $p_i$ from 
Eq. (\ref{pi}) it is  unnecessary for $E$ 
to be an eigenenergy. In fact, 
when $N$ is large enough, 
for any value of $E$ one can 
calculate a $f_+(p)$ distribution ($\gamma (E)$ is 
positive). On the other hand, for 
$p_i' = 1/p_{N-i+1}$ one can get a sequence 
of $p_i'$ corresponding to another realization 
of the random numbers and calculate a corresponding 
$f_-(p)$ distribution. Numerically, it has 
been found that when $N$ is large enough 
the form of the 
$f(p)$ distributions thus obtained 
is not sensitive to the value of $E$, that is, 
whether $E$ being an eigenenergy or not does not 
have any special influence on the form of $f(p)$. 
 Therefore, for the purpose of studying  
the shape of $f(p)$, one does not need to calculate 
 the exact eigenfunctions. 

  We have compared the above three methods numerically 
and found that they give the same results. 
Two comparisons  
between the second and third methods 
are given in Fig.2a 
and 2b, respectively, for the case of $v=1.0$ 
and $v=0.2$. In Fig.2a, the solid line represents 
$f_a(p)$ obtained by the third mothed for $E=-1.0$ 
($5 \times 10^5$ $p_i$ have been calculated)
and circles give $f_a(p)$ obtained by the second 
mothed of diagonalizing 30 Hamiltonian matrices 
with $N=1000$. For each result of the diagonalizations 
20 eigenfunctions with eigenenergies around $-1.0$
have been taken for calculating the $f_a(p)$. 
For the solid line and circles in Fig.2b, 
$E$ is about $-0.1$, 2000 Hamiltonian 
matrices with $N=250$ have been diagonalized 
and 10 eigenfunctions have been used for each 
matrix. The agreement between the two methods 
is quite clear in the figure. 
Since the third method is the most powerful and 
convenient, we will use it for the following 
calculations in this and the next sections.

\begin{center} { \bf B. Weak disorder } \end{center} 

    As  mentioned in section II, $f_+(p)$ and 
$f_-(p)$ must be different for the Anderson 
model, since  
$\gamma (E,\beta)$ for the increasing 
and decreasing parts have  different signs. 
However, in the case of weak disorder Lyapunov exponents 
$\gamma (E)$ are usually quite small. 
For example, when $v=4.0$, $\gamma (E 
\approx -4.0)$ is less than $1.0 \times 10^{-3}$. 
So it is possible for $f_+(p)$ to be close to 
$f_-(p)$.   Indeed, Fig.1a shows that 
they are quite similar. 

    As discussed in section II, the Lorentzian distribution 
in Eq. (\ref{lorentz}) is a fixed distribution for the mapping  
in Eq. (\ref{pi}) and when $\gamma (E)$ is 
small $f_{\pm}(p)$ 
may be close to $f_L(p)$. 
 Therefore, it is natural to fit the $f(p)$ 
distribution in 
Fig.1a by the Lorentzian distribution. Since $\overline 
\alpha _i$ in this case is $E/v \approx -1$, 
one can expect that parameters $b$ and $a$ 
for the Lorentzian distribution may be close to 
$\overline \alpha _i/2 \approx -0.5$ and $\sqrt{1-b^2} 
\approx 0.87$, respectively. Indeed, 
the best fit for the dots in Fig.1a 
by the Lorentzian distribution gives 
$b \approx -0.53$ and $a \approx 0.85$. 
The result is given in Fig.1b. 
It can be seen that the distribution $f_+(p)$ 
 is indeed quite close to the Lorentzian form. 

  For the other values of $E$ when $v=4.0$, 
it has been found that 
 when $|E|$ is larger than 0.2 
and within the region for eigenenergies 
(roughly speaking $|E| < 2v$), 
the shapes of $f_{\pm}(p)$ can also be 
fitted quite well by the same Lorentzian 
distribution as for $E \approx -4$. When $|E|$
becomes less than 0.2, the 
top of $f(p)$ will gradually deviate 
from that of the Lorentzian form, while tails 
are still of the Lorentzian form as  in Fig.1b. 
As an example, we present in Fig.3 
 central parts of $f_+(p)$ for  $E=0$ 
and its fitting curve of the Lorentzian form. For 
comparison, we have also given $f_+(p)$ 
distributions for $E=-4.0$, $E=-7.5$ 
and the corresponding fitting curves. The 
difference between $f_+(p)$ and $f_-(p)$ has also 
been studied for $0<|E|<0.2$ and 
has been found larger than that for the case 
of $|E|>0.2$. 
When $E=0$, $f_+(p)$ and $f_-(p)$ are almost 
the same. When $|E|$ becomes less than 0.2,
the form of $f_a(p)$ will also gradually deviate from 
that of the Lorentzian distribution in the top region. 

    When the value of $v$ is increased, 
 similar numerical results have also been obtained
for $|E|<2v$. 
Particularly, in the case of $v$ going to infinity
(or, equivalently, letting $w$ be zero while keeping 
$v$ a constant), the shape of $f(p)$ can also be fitted 
quite well by the Lorentzian distribution except  
for some special values of $E/v$, e.g., 0 or $-1$, 
For these values of $E/v$ some $p_i$ are zero
and the mapping in Eq. (\ref{pi}) can not procceed. But these 
special values are 
 unimportant because they do not correspond to any 
eigensolution of the Hamiltonian matrix.  
Since when $w=0$ 
eigenfunctions are extended and the Lyapunov 
exponents are zero,  the form of $f(p)$ 
may be exactly the Lorentzian form when $N$ goes to infinity. 

   When $v$ is decreased from 4, 
deviation of $f_{\pm}(p)$ from the Lorentzian distribution 
will also appear in the region of large $|E|$. 
As $v$ becomes smaller the region of deviation 
will become larger.  
For example, when $v=2$  for most of the values 
of  $|E|<2v$ the shape of $f_{\pm}(p)$ can be fitted quite 
well by the Lorentzian form, but when $v=1$ 
deviation of $f_{\pm}(p)$ from 
the Lorentzian form has been 
detected in the top region in most cases. 
Correspondingly when $v=1$ obvious difference 
between $f_+(p)$ and $f_-(p)$ has also been found. 
Since $f_+(p)$ and $f_-(p)$ can compensate 
each other, $f_a(p)$ may be of a better Lorentzian 
form than $f_{\pm}(p)$. In deed, 
when $v=1$ and $1.7 > |E| > 0.5$, 
the form of $f_a(p)$  has been found of  
a good Lorentzian form still in the top region.

  The above phenomena for relatively small 
values of $v$ can be explained by some results 
of the Appendix C. As shown there, an eigenfunction can be 
divided into perturbative and non-perturbative parts. 
For the perturbative part, Eqs. (\ref{cin2}) and (\ref{cde2}) 
tells that $p_i$ behaves differently for the increasing 
and decreasing parts of the eigenfunction. Also as shown 
in Appendix C, when 
$v$ becomes smaller the perturbative part will become 
larger, while for a fixed $v$ eigenfunctions with 
large $|E|$ have larger perturbative parts 
than those with relatively small $|E|$. As a result, 
 the difference between $f_+(p)$ and $f_-(p)$ 
will enlarge as $v$ becomes smaller
or $|E|$ becomes larger. 
 Since some dynamical effects remain in 
$A_{\alpha }( j \to j+1)$, $p_i$ of perturbative 
parts will also make $f_+(p)$ and $f_-(p)$ deviate 
from the Lorentzian distribution
(cf. the next subsection).

\begin{center} {\bf C. Strong disorder}  \end{center}  

    In the case of strong 
disorder (for small values of $v$), considerable difference 
between $f_+(p)$ and $f_-(p)$ has been found. An example is 
given in Fig.4 for $v=0.2$ and $E=-0.1$, where the
circles and the dashed line represent $f_+(p)$ and 
$f_-(p)$, respectively. Here $f_+(p)$ and $f_-(p)$ 
are distributions of $p_i$ and $1/p_i$, respectively, 
obtained from Eq. (\ref{pi}). 
As discussed in section II and 
Appendix B, long tails of $f(p)$ 
should  behave as $1/p^2$. In Fig.5a and 5b we show the 
above $f_+(p)$ and $f_-(p)$ distributions (dots) in logarithm
scale and the corresponding fitting curves (solid lines)
 of the Lorentzian form for their tails. 
For the $f_+(p)$, due to the form of the central part, 
 we can fit only 
one of its two tails by the Lorentzian distribution once. 
For the other tail similar fitting can also be done. 
It is quite clear from the figure that long tails of both 
$f_+(p)$
and $f_-(p)$ can be fitted well by the Lorentzian 
distribution. 

    In Fig.4 it can be seen that 
$f_+(p)$ changes only a little when $p$ varies 
from $-2$ to 1. This can be explained 
by perturbation theory. In fact, 
 Eq. (\ref{cin2}) in Appendix C shows that $p_i$ for the 
perturbative part of an eigenfunction  satisfies 
\begin{equation} p_{j} \approx \frac 1{A_{\alpha }(j-1 \to j)} 
\label{pj+1} \end{equation} 
where, according to Eq. (\ref{Anear}), 
\begin{equation} A_{\alpha } (j-1 \to j) = 
\frac v{E - \epsilon_{j-1}} + O(v^3). 
\label{Aa} \end{equation} 
To the first order of approximation, $p_j \approx  
(E- \epsilon_{j-1})/v$. So when $v$ is small, the
contribution of perturbative parts tends to make $f_+(p)$ 
close to a platform in the region of $p \in [(E-0.5)/v, 
(E+0.5)/v]$. Since there also exist non-perturbative 
parts for which Eq. (\ref{pj+1}) does not hold and 
when $v$ is not quite small higher order of approximation 
should also be considered, the region of the platform 
should be narrower. This can be seen quite clearly 
in Fig.6, in which $f_+(p)$ and $f_-(p)$ 
for $v=0.1$ and $E=0.1$ are presented. 

\begin{center} {\Large IV. Two other versions of 1D 
tight binding model } \end{center}

\begin{center} {\bf A. Paired correlated Anderson model  }
\end{center} 

    When some correlation is
introduced into the diagonal elements of the Hamiltonian 
matrix of the Anderson model, delocalization can appear. 
For example, one can arrange $\epsilon_i$ into pairs with 
$\epsilon _{2k} = \epsilon_{2k+1}$, where $\epsilon_{2k}$ 
 for different pairs are random numbers \cite{Heinrichs}. 
In some energy regions, localization lengths for 
this model can be far larger than that for the corresponding 
standard Anderson model (for some special energies 
localization lengths may be infinite). 
But for local statistical properties 
as the shape of $f(p)$, it is reasonable to expect that the two 
models  may show 
similar behaviors. In deed, similar properties 
as shown in section III for $f(p)$ of the Anderson 
model has also been found numerically for $f(p)$ 
of this model (Fig.7).
Particularly, in the case of weak disorder, 
central parts of $f(p)$ for $|E|$ 
around zero can be fitted  by the Lorentzian 
form better than for the corresponding case of the 
standard Anderson model (see Fig.7b compared with Fig.3c). 

\begin{center} {\bf B. 1D quasi-periodic tight 
binding model } \end{center} 

    For this model, we take $\epsilon_i$ in 
Eq. (\ref{Hele}) as 
\begin{equation} \epsilon_i = cos(2 \pi i \sigma )/2
\label{harp} \end{equation} 
where $\sigma$ is an irrational number. 
Numerically it has been found that 
when $v$ is large, e.g., $v \ge 4$, the form of $f(p)$ can 
by fitted quite well by the Lorentzian form 
(Fig.8a) when $E \ne 0$
and  $f_+(p)$ is almost the same as $f_-(p)$ 
since eigenfunctions are extended. 
Even for quite small values of $|E|$, the fitting is 
also quite good. But when $E=0$, $f(p)$ 
has a high peak in the neighbourhood of $p=0$. 
Similar to the Anderson model, when $v$ is 
decreased from 4, the form of $f(p)$ will deviate 
gradually from that of the Lorentzian distribution,
especially for eigenfunctions with 
relatively large $|E|$. This is also due to the fact that for 
these eigenfunctions the perturbative parts become 
larger as $v$ becomes smaller. 
In some cases it has been found that although $f_+(p)$ 
and $f_-(p)$ deviate from the Lorentzian form notably 
they are quite close to each other. 
 When $v$ is smaller than 1, many localized 
states can be found. For most of these localized 
states $f_+(p)$ have shapes much more irregular than that for 
the Anderson model. 
In Fig.8b we give an example for $v=0.2$ and $E=-0.1$. 
Circles represent $f_+(p)$ and the 
solid line is  the fitting 
curve of the Lorentzian form for the far right tail. The far 
left tail and the middle right tail (for $p$ between 
3 and 7) can also be fitted by the Lorentzian 
form but with different fitting parameters. 

\begin{center} {\Large V. Conclusions and discussions}
\end{center} 

The distribution $f(p)$ of ratio of nearest components 
of energy eigenfunctions, 
$p_i=c_i/c_{i-1}$,
  is studied numerically 
in this paper for three versions of the 1-dimensional tight
binding model: the Anderson model, a quasi-periodic 
tight binding model and 
a paired correlated Anderson model. 
It is shown that $f(p)$ for increasing and decreasing 
parts of eigenfunctions 
(denoted as $f_+(p)$ and $f_-(p)$, respectively)
should be treated separately. Numerically the following 
results have been found for all the three models. (1)
In the case of strong perturbation (weak disorder), 
when $|E|$ is not close to 
zero, $f_+(p)$ and $f_-(p)$ are close to each other and can be 
fitted quite well by the Lorentzian distribution. 
(2) In the case of weak perturbation 
(strong disorder), $f_+(p)$ deviates notably
 from $f_-(p)$ and only their tails can be fitted well 
by the Lorentzian distribution. Some features of the shape of 
$f_+(p)$ in the case of strong disorder for the Anderson 
model can be explained by perturbation theory. 

Eq. (\ref{pi}) shows that once an energy $E$ is given, 
$p_1, p_2, \cdots p_N$ can be calculated readily 
from the $i=0,1, \cdots N-1$ cases of the equation, despite of  
whether $E$ is  an eigenenergy or not (Eq. (\ref{pi}) 
for $i=N$ is the condition for $E$ to be an eigenenergy). 
Numerically it has been found that $E$ being an eigenenergy or not 
has no influence on whether $f(p)$ can be fitted well by the
Lorentzian form. That is to say, the closeness between 
the distribution $f(p)$ and the Lorentzian 
distribution in the case of weak 
disorder is not specific for eigensolutions. It is in fact 
determined by the form of the mapping in Eq. (\ref{pi}). 
We would like to point out 
that this property of $f(p)$ is, in fact, 
related to a more general problem, that is, 
some information for properties 
of eigenfunctions may be obtained without diagonalizing the 
Hamiltonian matrix. 

\begin{center} {\Large Acknowledgments } \end{center}

The authors are very grateful to 
Bao-wen Li, Guang-shan Tian 
and Pei-qing Tong for valuable discussions. 
The authors are also thankful to J.M. Luck for 
introducing us to some previous 
related work on the Riccati variable. 
This work was partly supported by grants from the Hong 
Kong Research Grant Council (RGC), the Hong Kong Baptist 
University Faculty Research Grant (FRG), 
Natural Science Foundation of China
and 
the National Basic Research Project ``Nonlinear 
Science'', China.

\appendix
\section{Properties of $A_i$ and $B_i$ }

    In order to express $A_i$ and $B_i$ explicitly
by making use of $\alpha _1, \cdots , \alpha _i$, 
we first introduce a quantity $S_{ji} (0 < j< i)$ 
defined by 
\begin{equation} S_{ji} = \alpha _j \alpha _{j+1} 
\cdots \alpha _i. \label{sji} \end{equation} 
Taking out $k$ pairs of neighbouring $\alpha _l$ 
arbitrarily from $S_{ji}$, one has a product of 
$(i-j+1-2k)$ $\alpha _{l'}$,  denoted 
as $S_{ji}^{(km)}$. 
Let us define another quantity $D_{ji}$ as 
\[ D_{ji} = S_{ji} - \sum_m S_{ji}^{(1m)}
+ \sum_m S_{ji}^{(2m)} - \cdots \]
\begin{equation} = S_{ji} + (-1)^k \sum_{(k,m)}
S_{ji}^{(km)}. \label{dji} \end{equation}
Then, making use of Eq. (\ref{pi}) one can proof that 
\[ A_i = D_{1i} \] 
\begin{equation} B_i = -D_{2i} 
\label{AB} \end{equation}

    The following relations can be readily obtained from 
Eqs. (\ref{pi}), (\ref{dji}) and (\ref{AB}), 
\begin{equation} p_i = \frac{D_{ji} p_j -D_{(j+1)i}}
{D_{j(i-1)}p_j - D_{(j+1)(i-1)}}, \label{pjpi} \end{equation} 
\begin{equation} p_j = \frac{D_{(j+1)i} -  p_iD_{(j+1)(i-1)}}
{D_{ji} - p_iD_{j(i-1)}}. \label{pipj} \end{equation} 
Eq. (\ref{p1pi}) is a special  case of Eq. (\ref{pjpi}). 
Conversely, $p_1$ can also be expressed as a function of $p_i$, 
\begin{equation} p_1 = \frac{p_i B_{i-1} - B_i}{A_i - p_i A_{i-1}}. 
\label{p1pn} \end{equation}
Making use of Eq. (\ref{p1pi}) and then (\ref{p1pn}), 
one has
\[ p_1 p_2 \cdots p_{n-1} = A_{n-1} p_1 + B_{n-1} \] 
\begin{equation} = \frac{A_nB_{n-1} - A_{n-1}B_n}
{A_n - p_n A_{n-1}}. \label{prod2} \end{equation}
On the other hand, from Eq. (\ref{pipj}) for $i=n$ 
and $j=1,2,\cdots , n-1$, the following relation 
can be obtained readily, 
\begin{equation} p_1p_2 \cdots p_{n-1} = 
\frac 1{A_n - p_n A_{n-1}}. \label{prod} \end{equation}
Then, substituting Eq. (\ref{prod})  into  (\ref{prod2})
one has 
\begin{equation} A_n B_{n-1} - A_{n-1} B_n =1.  
\label{AnBn} \end{equation}
So for large $|A_n|$ and $|B_n|$, 
$A_n/B_n$ is approximately a constant. 

    An interesting result of Eqs. (\ref{AnBn}) and 
(\ref{p1pi}) for the 
case of large $|A_n|$  is  that different $p_1$ gives 
approximately the same $p_n$. This means  
 if $|A_n|$ approaches infinity when $n$ goes to infinity, the 
distributions $f(p)$ for sequences start from different $p_1$
must be the same. 

    In principle, once  $A_i$ and $B_i$ 
are known,  explicit expressions for 
properties of the corresponding eigenfunction 
can be obtained. For example, 
    from Eqs. (\ref{Lyap}) and (\ref{p1pi}), it can be  
seen that the Lyapunov  exponent is determined by properties 
of $A_n$ and $B_n$, 
\begin{equation} \gamma (E) = \lim_{n \to \infty} 
\frac 1n \ln |A_n + \frac{B_n}{p_1}|. 
\label{LyapuAB} \end{equation}
However, to obtain practically useful expressions 
for $A_i$ and $B_i$ from Eqs. 
(\ref{dji}) and (\ref{AB}) is mathematically 
quite difficult.

\section{$F_n(p)$ and $f_L(p)$}

    For a set of  $p_1$ with some distribution $F_1(p_1)$, 
from probability theory (see, e.g., 
\cite{Beaumont}) and Eq. (\ref{pi}) it is easy to find out the 
relation between $F_n(p_n)$ and $F_1(p_1)$
\begin{equation} F_n(p_n) = (p_1p_2 \cdots p_{n-1})^2 
F_1(p_1) |_{p_n} \label{FnF1} \end{equation}
where $p_1, p_2, \cdots, p_{n-1}$ on the right hand side  are 
functions of $p_n$. 
From Eqs. (\ref{p1pn}) and (\ref{prod}), it can be 
found out that 
\begin{equation} F_n(p) = \frac 1{(A_n - p A_{n-1})^2} 
F_1(\frac{pB_{n-1} - B_n}{A_n - pA_{n-1}}) 
\label{Fn2} \end{equation}
where for simplicity the subscript of $p_n$ has been omitted. 
Due to the fact that $\alpha _i$ is not a constant, 
$F_n(p)$ can not approach a fixed distribution as $n$ goes 
to infinity.

The sum of $F_n(p)$, 
\begin{equation} F(p,n) \equiv \frac 1n \sum_{i=1}^n F_i(p), 
\label{Fpn} \end{equation} 
is more important than $F_n(p)$, since it 
  is just  the average of the 
distributions $f(p)$ for $p_i$ obtained from the 
set of $p_1$ by Eq. (\ref{pi}). 
Numerically it has been found that 
when $f(p)$ is close to $f_L(p)$ in Eq. (\ref{lorentz}),
$F(p,n)$ first approaches 
 $f_L(p)$ when $n$ increases from $n=1$, despite of 
which form $F_1(p_1)$ is. However, when $n$ becomes larger 
than some quantities $F(p,n)$ will show quite large fluctuations 
if the Lyapunov exponent is positive. This is not difficult 
 to understand since positive Lyapunov exponent makes 
$|A_n|$ increases exponentially and Eq. (\ref{Fn2}) shows 
that $F_n(p)$ has peaks in a neighborhood of $q_n=A_n/A_{n-1}$
which becomes smaller as $|A_n|$ increases. 
    Equation (\ref{Fn2}) also shows that 
the form of $F_1$ has considerable influence on 
the form of $F_n(p)$ only in the neighborhood of $q_n$. 
Out of the neighborhood, $F_n(p)$ behaves like $1/(p-q_n)^2$. 
 Therefore, $F(p,n)$ (so $f(p)$) should  
behave as $1/p^2$ in the tail region. 

    In order to calculate the value of $\gamma = 
\int f_L(p) ln|p| dp$ for $f_L(p)$ in Eq. (\ref{lorentz}),
 we first change the variable 
to $q=\alpha _i -1/p$, which gives 
\begin{equation} \gamma = \int ^{\infty }_{-\infty }
\frac{a/\pi}{(q-b)^2 +a^2} ln |\frac 1{\alpha _i -q}| 
dq. \end{equation} 
Then, changing $q$ to $x=2b-q$ we have 
\begin{equation} \gamma =- \int ^{\infty }_{-\infty }
\frac{a/\pi}{(x+b)^2 +a^2} ln |x| 
dx. \end{equation} 
Finally, after changing $x$ to $y=-x$ 
one can see that $\gamma =-\gamma$ which gives $\gamma =0$.

\section{Perturbative study for 1D Anderson model}

    In this appendix we study 1D Anderson model by 
making use of 
a generalization of the so-called Brillouin-Wigner perturbation 
expansion (GBWPE) \cite{WIC97}.  
According to GBWPE, when perturbation is not extremely strong, each 
eigenfunction  can be divided into two parts, one of which  
 can be expressed as a convergent perturbation expansion
by making use of the other. For the former part, an 
expression for $p_j$ will be given
in this appendix, which is useful for 
understanding some features of the shape of $f(p)$ 
when perturbation is not strong. 

    For the 1D tight-binding 
model with Hamiltonian $H = H^0+V$, 
we denote eigenstates of $H^0$ as $|k>$ 
and eigenstates of $H$ as $|\alpha >$, 
\begin{equation} H^0 |k> = E^0_k |k> \end{equation} 
\begin{equation} H |\alpha > = E_{\alpha } |\alpha >.  
\label{HE}  \end{equation} 
($E^0_k=\epsilon_k$ in Eq. (\ref{Hele}))
    For the sake of completeness, we first give 
a brief  discussion for
 the above mentioned GBWPE. 
Dividing the 
set of basis states $|k>$ 
 into two parts denoted by $S_P$ (including $N_P$ basis states) 
and $S_Q$ (including $N_Q$ basis states), respectively,  
one has two projection operators 
\begin{equation} P \equiv \sum_{|k> \in S_p} |k><k|, 
\ \ \ \ \ Q \equiv 1-P. 
\label{PQ} \end{equation}
Subspaces related to $P$ and $Q$ will be called 
in the following the {\it P} and 
{\it Q subspaces }, respectively.
 Splitting an arbitrary eigenstate $|\alpha>$
 into two orthogonal parts
$ |\alpha> = 
|t> + |h> $ 
where $|t> \equiv P|\alpha>$ and $|h> \equiv Q |\alpha>$, 
it can be shown, by making use   of the stationary 
  Schr\"{o}dinger equation, that 
\begin{equation} 
|\alpha>= |t> + \frac 1{E_{\alpha} - H^0} 
QV|\alpha>.  
\label{GBW} \end{equation}
The iterative expansion of Eq. (\ref{GBW}) gives  
\[  |\alpha> = |t> +
\frac 1{E_{\alpha}-H^0}Q V|t>  
+ (\frac 1{E_{\alpha}-H^0}Q V)^2|t>  \] 
\begin{equation} 
\ \ \ \ \ \ \ + (\frac 1{E_{\alpha}-H^0}Q V)^3|t> 
 + \cdots
 \label{e-psi} \end{equation}
if 
\begin{equation} \lim _{n \to \infty } <T^n_{\alpha }|T^n_{\alpha }> 
=0  \label{Tn} \end{equation}
where 
\begin{equation} 
|T_{\alpha}^n> \equiv 
 (\frac 1{E_{\alpha}-H^0}Q V)^n|\alpha>.
\end{equation}
 Here the eigenvalue $E_{\alpha }$ has 
been treated as a constant. Eq. (\ref{e-psi}) is just the above mentioned 
 {\it generalization  of }
{\it    Brillouin-Wigner perturbation 
expansion (GBWPE)}. 
For convenience, in what follows let us use the following 
notations:  $|i>$ denoting a basis state 
in the $P$ subspace and $|j>$ denoting a basis state in the $Q$
subspace. Then, Eq. (\ref{e-psi}) gives 
\[ C_{\alpha j} \equiv <j|\alpha> = 
\sum_{i \in P} ( \frac { V_{ji}}{E_{\alpha} - E^0_j} 
+ \sum_{k \in Q} \frac { V_{jk}}{E_{\alpha} - E^0_j} 
\frac { V_{ki}}{E_{\alpha} - E^0_k} \] 
\begin{equation} 
+ \sum_{k,l \in Q} \frac { V_{jk}}{E_{\alpha} - E^0_j} 
\frac { V_{kl}}{E_{\alpha} - E^0_k} 
\frac { V_{li}}{E_{\alpha} - E^0_l} \cdots )
C_{\alpha i}
 \label{cja} \end{equation}

    Generally to say, $<T_{\alpha }^n|T_{\alpha}^n>$  vanishes 
as $n \to \infty$  
if $P$ and $Q$ subspaces are such chosen that, for 
any  state $|j>$ in the $Q$ subspace, $|E_{\alpha} -E^0_j|$ is large 
enough compared with $ V$, i.e., if all the 
basis states $|k>$ with  
small $|E_{\alpha }- E^0_k|$ are in the $P$ subspace. 
So  there are generally many $P$ (and correspondingly $Q$) 
subspaces 
 ensuring the validity of Eq. (\ref{e-psi}). 
In these $P$  subspaces 
there exists one with the minimum number of basis states. 
   Projection operators related to this $P$ subspace and the 
corresponding $Q$ subspace are denoted by $P_{m }$ 
and $Q_{m }$, respectively. 
$|t_{\alpha }> \equiv P_{m }|\alpha>$ 
will be called in the following 
 the {\it non-perturbative part} of $|\alpha >$, 
and  $|h_{\alpha }> \equiv Q_{m }|\alpha>$ 
the {\it perturbative part}, 
which can be expressed in terms of $|t_{\alpha}>$, $E_{\alpha }$, 
$ V$ and $H^0$ as shown in Eq. (\ref{e-psi}). 
In what follows, we will discuss  the $P_m$ and $Q_m$ subspaces 
only and for the sake of brevity we will omit the subscript $m$. 
It is clear from the above discussion that the subspace $P$ 
will increase (correspondingly, $Q$ will decrease) as the 
perturbation $v$ becomes stronger. It can also be seen that 
for the Anderson model eigenfunctions with larger 
$|E - \overline {\epsilon_i}|$ 
have larger perturbative parts.

    In order to  study $C_{\alpha j}$ of 
the perturbative part of $|\alpha >$ in Eq. (\ref{cja}),  
we make use of the concept of path, in analogy to that in  the 
  Feynman's path integral theory \cite{F65}. 
For $(n+1)$ basis states $|k_0>, |k_1>, \cdots |k_{n-1}>, 
|k_n>$ with $|k_0>, \cdots |k_{n-1}>$ in the $Q$ subspace and 
$|k_n>$  in either the $Q$ or the $P$ subspace, 
 we  term the sequence  $k_0 \to k_1 \to 
\cdots \to k_{n-1} \to k_n$ 
{\it a path of n paces} from $k_0$ to $k_n$,
if $V_{kk'}$ corresponding to  each pace is non-zero. 
Clearly, paths from $k_0$ to 
$k_n$ are determined by the structure of the Hamiltonian matrix in 
the $H^0$ representation. 
Attributing a  factor $V_{kk'}/(E_{\alpha} -E^0_k)$ to each 
pace $k \to k'$, we define  
 the contribution of a path $s$ (from $j$ to $i$)  to $C_{\alpha j}$, 
denoted by $f_{\alpha s}(j \to i)$,   
 as the product of  the  factors of all its  paces.  
  Then,  $C_{\alpha j}$ in Eq. (\ref{cja})  
 can be rewritten as
\begin{equation} C_{\alpha j} = \sum _{i \in P} A_{\alpha } (j \to i)  
C_{\alpha i}  \label{CA} \end{equation}
where 
\begin{equation}
A_{\alpha } (j \to i) =  \sum_s f_{\alpha s}
(j \to i)  
\label{Af} \end{equation}
and $s$ denotes  possible paths from $j$ to $i$. 

    An interesting property of $C_{\alpha j}$ given by Eq. 
(\ref{cja}) is that they satisfy equations 
\begin{equation} \sum _k H_{jk} C_{\alpha k} = E_{\alpha } 
C_{\alpha j}   \label{he} \end{equation} 
even for arbitrary values of $E_{\alpha }$ and $C_{\alpha i}$. 
In fact, from the definition of $A_{\alpha }
(j \to i)$  in Eq. (\ref{Af}), it can be shown that  
\begin{equation} \displaystyle  A_{\alpha }(j \to i) = 
\sum _{j' \ne j} \frac{V_{jj'}}{E_{\alpha }-E^0_j} 
A_{\alpha }(j' \to i) + 
\frac{V_{ji}}{E_{\alpha }-E^0_j}. 
\label{AA'}  \end{equation} 
Substituting Eq. (\ref{AA'}) into Eq. (\ref{CA}), one has 
\begin{equation} \displaystyle  C_{\alpha j} = 
\sum _{j' \ne j} \frac{V_{jj'}}{E_{\alpha }-E^0_j} 
C_{\alpha j'} + \sum_i \frac{V_{ji}}{E_{\alpha }-E^0_j}C_{\alpha i},  
\label{CC'}  \end{equation} 
which gives Eq. (\ref{he}). Therefore,  the values of $E_{\alpha }$ 
and $C_{\alpha i}$ must be determined by the other part of the 
eigen-equation (\ref{HE}), i.e., by equations 
for $C_{\alpha 
i}$, \begin{equation} \sum _k H_{ik} C_{\alpha k} = E_{\alpha } 
C_{\alpha i},    \label{Ht} \end{equation} 
and the normalization condition (except for a common 
phase). 

    Now let us apply the above results to the study of the 
Hamiltonian in Eq. (\ref{Hele}). 
 Firstly, we discuss the case in which the value of $v$ is 
so small that there is only one basis state $|k>$ satisfying 
$|E_{\alpha } - E^0_k| < v/2$. (For such a small perturbation strength 
Brillouin-Wigner  perturbation expansion 
is also convergent.) In this 
case, the $P$ subspace corresponding to the non-perturbative 
part of $|\alpha >$ is just this basis state $|k>$. Let us 
denote it by $|i>$. Then, for each $|j>$ in the $Q$ subspace, 
according to Eq. (\ref{CA}) 
\begin{equation}  C_{\alpha j} = A_{\alpha }( j \to i) 
C_{\alpha i} . \label{CA1}  \end{equation} 
Due to the specific form of perturbation as shown  in 
Eq. (\ref{Hele}), paths starting from $j < i$ can never reach 
points $k >i$ and {\it vice versa}, 
i.e., paths can not cross ``$i$''. 
From the discussion of Eqs. (\ref{he}) and (\ref{Ht}), 
we know that $C_{\alpha j}$ given by Eq. (\ref{CA1}) 
satisfy Eq. (\ref{he}) naturally and the value of $E_{\alpha }$ 
is determined by the equation 
\begin{equation} v C_{\alpha i-1} + E^0_i C_{\alpha i} 
+ v C_{\alpha i+1} = E_{\alpha } C_{\alpha i} 
\label{Ht1}  \end{equation}
where $C_{\alpha i-1}$ and $C_{\alpha i+1}$, given 
by paths left to and right to $i$, respectively, are functions 
of $E_{\alpha }$ (see Eqs. (\ref{CA1}) and (\ref{Af})). 

    For simplicity, let us first discuss the case 
$j < i$. According to the definitions of $A_{\alpha }(j \to i)$ 
and $f_{\alpha s}(j \to i)$, one can find out that  
\begin{equation}  A_{\alpha }(j \to i) = A_{\alpha }
(j \to j+1) A_{\alpha }(j+1 \to i), 
\label{AA1}   \end{equation} 
where 
\[ A_{\alpha }(j \to j+1) = 
f_{\alpha s_1}(j \to j+1) 
+f_{\alpha s_2}(j \to j+1) \]
\begin{equation}  
+f_{\alpha s_3}(j \to j+1)  
+f_{\alpha s_4}(j \to j+1) 
+ O(v^5).  \label{Anear}  \end{equation} 
Paths $s_1, s_2, s_3$ and $s_4$ in Eq. (\ref{Anear}) are 
$s_1: j \to j+1$, $s_2: j \to j+1 \to j \to j+1$, 
$s_3: j \to j+1 \to j+2 \to j+1$, and  
$s_4: j \to j-1 \to j \to j+1$, respectively, and $O(v^5)$ 
represents the contribution of paths with more than 3 
paces.  
Substituting Eq. (\ref{AA1}) into (\ref{CA1}), one has  
\begin{equation} p_{j+1} = \frac{C_{\alpha j+1}}
{C_{\alpha j}} =  \frac 1{A_{\alpha }(j \to j+1)}; 
 \label{cin} \end{equation} 
and
\[ C_{\alpha j} = 
A_{\alpha }(j \to j+1) A_{\alpha }(j+1 \to j+2) \] 
\begin{equation} 
\cdots  
A_{\alpha }(i-1 \to i) C_{\alpha i}. 
\label{CCA2} \end{equation} 
 Thus, on average, 
\begin{equation} C_{\alpha j} \approx (\overline A)^{i-j} 
C_{\alpha i} \label{final1}  \end{equation} 
where $\overline A$ is the average  value of $A_{\alpha }(j \to j+1)$. 
From Eq. (\ref{Anear}) one can  see that $|\overline A|<1$, 
so $|C_{\alpha j}|$ increases exponentially as $j$ 
approaches $i$. 
When $j > i$, similar arguments lead to the following 
relations 
\begin{equation}  A_{\alpha }(j \to i) = A_{\alpha }
(j \to j-1) A_{\alpha }(j-1 \to i), 
\label{AAp}   \end{equation} 
\begin{equation} p_{j} = \frac{C_{\alpha j}}
{C_{\alpha j-1}} =  A_{\alpha }(j \to j-1) 
 \label{cde} \end{equation} 
and 
\begin{equation} C_{\alpha j} \approx (\overline A)^{j-i}
C_{\alpha i}. \label{finaln}  \end{equation}
 
    Secondly, we discuss the case of $v$ being larger 
than only a few $|E_{\alpha }  -E^0_k|$. (This is a case in which 
Brillouin-Wigner perturbation expansions diverge.) In this case, 
the $P$ subspace corresponding
to the non-perturbative part of $|\alpha >$ is                 
composed of more than one basis states with $E_i^0$  
close to $E_{\alpha }$ distributed randomly in 
the region $[-w/2,w/2]$ . Since 
$N_p$, the number of basis states in the $P$ subspace,   
 is much smaller than $N$, the total number of basis states, 
labels $i$ are generally separated 
by labels $j$ of $|j>$ in the $Q$ subspace  and  
there is generally no successive labels of $i$. 
On the other hand, $S_Q$, the set of $|j>$,  is also  
separated into a series of subsets by $|i>$. 
 For convenience, we denote the basis states $|i>$ in 
increasing order by $|i^{(1)}>, |i^{(2)}>, \cdots |i^{(q)}>, 
\cdots |i^{(N_P)}>$ with $i^{(1)}< i^{(2)}< \cdots < i^{(q)} 
\cdots < i^{(N_P)}$ and denote the corresponding  
 series of  subsets of $S_Q$ by 
$S^{(1)}_Q, S^{(2)}_Q, \cdots S^{(q)}_Q, \cdots $. 

    To apply GBWPE to this case of perturbation strength, 
let us consider a state $|j>$ in $S_Q^{(q)}$ satisfying 
$i^{(q)} < j < i^{(q+1)}$. Due to the same reason as in 
the first case, paths starting from $j$ can 
not cross $i^{(q)}$ and $i^{(q+1)}$, that is, they are 
restricted in the region $[i^{(q)}, i^{(q+1)}]$.  
Then, according to Eq. (\ref{CA}) 
\begin{equation}  
C_{\alpha j} = A_{\alpha }(j \to i^{(q)}) \cdot  
C_{\alpha i^{(q)}} + A_{\alpha }(j \to i^{(q+1)}) \cdot 
C_{\alpha i^{(q+1)}}. 
\label{CA2} \end{equation} 
Similar to Eq. (\ref{AA1}) in this case we have  
\[  A_{\alpha }(j \to i^{(q)}) = A_{\alpha }
(j \to j-1) A_{\alpha }(j-1 \to i^{(q)}) \]  
\begin{equation}    A_{\alpha }(j \to i^{(q+1)}) = A_{\alpha }
(j \to j+1) A_{\alpha }(j+1 \to i^{(q+1)}).   
\label{AA2}   \end{equation} 
Then, 
\[ C_{\alpha j} = 
A_{\alpha }(j \to j-1)  \cdots  
A_{\alpha }(i^{(q)}+1 \to i^{(q)}) C_{\alpha i^{(q)}} \] 
\begin{equation}  + 
A_{\alpha }(j \to j+1)  \cdots  
A_{\alpha }(i^{(q+1)}-1 \to i^{(q+1)}) C_{\alpha i^{(q+1)}}  
\label{CCA3} \end{equation} 
\[ \approx (\overline A)^{j-i^{(q)}} C_{\alpha i^{(q)}} 
+ (\overline A)^{i^{(q+1)}-j} C_{\alpha i^{(q+1)}} \] 
Since only a few $|E_{\alpha }-E^0_i|$ for $|i>$ in the 
$P$ subspace are smaller than 
$v$, Eq. (\ref{Anear}) also gives $|\overline A|<1$. 

    There are $N_P$ equations as Eq. (\ref{Ht}) which 
determine the value of $E_{\alpha }$ and give $(N_P-1)$ 
relations between $C_{\alpha i}$. What is of special importance 
here is that $E_{\alpha }$ is determined mainly by one 
of these equations only. In fact, 
suppose $|C_{\alpha i^{(m)}}|$ is the largest one among 
$|C_{\alpha i}|$. For $j_m=i^{(m)}-1$ and $j_m'=i^{(m)}+1$, 
Eq. (\ref{CCA3}) shows that both $C_{\alpha j_m}$ 
and $C_{\alpha j_m'}$ are mainly given by terms including 
$C_{\alpha i^{(m)}}$. From the discussions for Eqs. 
(\ref{Ht}) and (\ref{Ht1}), one can see that 
 $E_{\alpha }$ is mainly determined by equation, 
\begin{equation} 
 E_{\alpha }- E^0_{i^{(m)}}  \approx 
vA_{\alpha }(j \to i^{(m)})
+ vA_{\alpha }(j' \to i^{(m)}), 
\label{EE1} \end{equation} 
i.e., by  contribution of paths around 
$i^{(m)}$. 

    As a result, for any other state $|i>$ with 
$i = i^{(q)} \ne i^{(m)}$, it is generally  
impossible for both $C_{\alpha j}$ and $C_{\alpha j'}$ 
($j=i-1$ and $j'=i+1$) to be 
mainly determined by terms including $C_{\alpha i}$. In fact, 
if that is possible, one will have an equation similar to Eq. 
(\ref{EE1}), 
which means that $E_{\alpha }$ is mainly determined by the contribution 
of paths around $i$. 
Generally, to say, this contradicts with Eq. (\ref{EE1}). 

Then, at least one of $C_{\alpha j}$ and $C_{\alpha 
j'}$   should have a  considerable contribution 
from the other $C_{\alpha i'}$ ($i'=i^{(q-1)}$ or 
$i'=i^{(q+1)}$). 
For example, Suppose $C_{\alpha j}$ 
has a considerable contribution from  $C_{\alpha i'}$ with 
$i'=i^{(q-1)}$.  
This means   $C_{\alpha j}$ is in the 
same order of magnitude as 
\[ (\overline A)^{(i-i'-1)} C_{\alpha i'}  \] 
(see Eq. (\ref{CCA3})). 
Since according to the eigen-equation (\ref{he}) 
\begin{equation} v C_{\alpha i} = (E_{\alpha } - E^0_j) 
C_{\alpha j} - v C_{\alpha j-1}, \end{equation}
one can see that 
$C_{\alpha i}$ is nearly in the same order of magnitude as 
$C_{\alpha j}$. Thus, for $j_b$ satisfying  $i>j_b>i'$ 
Eq. (\ref{CCA3}) gives 
\begin{equation} C_{\alpha j_b} \approx (\overline A)^{j_b-i'} 
C_{\alpha i'}. \label{CAC2}  \end{equation} 
Similar result can also be obtained if $C_{\alpha j'}$ 
has a considerable contribution from $C_{\alpha i^{(q+1)}}$. 
Therefore,  when $k$ changes from $i^{(q)}$ to 
$i^{(q+1)}$ or from $i^{(q)}$ to $i^{(q-1)}$, $|C_{\alpha k}|$ 
should on average increase exponentially in at least one of 
the two cases.  Specifically, if $i^{(q)}=1$ for $q=1$, $|C_{\alpha k}|$ 
should on average increase exponentially when $k$ increases   
 from $i^{(1)}$ to $i^{(2)}$. Similarly, for the case of 
$i^{(N_p)} =N$, $|C_{\alpha k}| $ should  increase 
exponentially when $k$ decreases from $N$ to $i^{(N_p-1)}$. 

    Now it is easy to show that the eigenfunction of $|\alpha >$ 
should be localized. Let us 
start from the basis state $|1>$. If $|1>$ is in the $P$ 
subspace, as discussed above, $|C_{\alpha k}|$ should on 
average  increase exponentially when $k$ changes from 
$1$ to $i^{(2)}$. If $|1>$ is in the $Q$ subspace, Eq. 
(\ref{CCA3}) shows that $|C_{\alpha k}|$ also on average 
increases exponentially until $k$ reaches the first 
$i=i^{(1)}$. 
Then, making use of the result of the above paragraph, 
it can be found that
 $|C_{\alpha k}|$ should  increase 
exponentially as $k$ changes from 1 to $i^{(m)}$. 
Similarly, it is easy to see  
that $|C_{\alpha k}|$ should  on average decrease 
exponentially when $k$ changes from $i^{(m)}$ to $N$. 
Thus, the eigenfunction of $|\alpha >$ has the 
property of exponential decay. 
Applying this result to Eqs. (\ref{CA2}) and (\ref{AA2}), one 
can found out the following relations 
\begin{equation} p_{j} \approx 
\frac 1{A_{\alpha }(j-1 \to j)}
 \label{cin2} \end{equation}
and 
\begin{equation} p_{j} \approx 
A_{\alpha }(j \to j-1)
 \label{cde2} \end{equation}
for the increasing and decreasing parts of eigenfunctions, 
respectively. 

    Thirdly, let us increase the value of $v$ further. 
In this case, there will be some successive labels $i$. That is 
to say, some of the subsets of $S_P$  
 separated by $S_Q$   
 may have more than one basis states $|i>$. When the 
numbers of basis states $|i>$ in the subsets of $S_P$ 
 are small compared with the numbers of basis states $|j>$ in the subsets 
of $S_Q$, following similar 
arguments as in the second step case one can see that 
when $|\overline A|<1$ energy 
eigenfunctions should  also  be localized (exponential decay). 
Finally, when the number of basis states in $S_P$ is larger 
than that in $S_Q$, 
exponential decay of eigenfunctions  can not be proved by 
the method used above, but can be proved by other methods
(\cite{Ishii}). 

    However, in these two cases 
the method is still useful for studying properties 
of $p_j$ and two results can be obtained. 
 First, for perturbative parts in a region of $j \in 
[j_1,j_2]$, if one of the values of $|C_{\alpha j_1}|$ 
and $|C_{\alpha j_2}|$ is far larger than the other, 
 (\ref{cin2}) and (\ref{cde2}) can still 
be obtained from equations  similar to Eqs. (\ref{CA2}) and 
(\ref{AA2}) 
 Second, for $|E_{\alpha}|> w/2$, since 
larger $|E_{\alpha }|$ means larger $|E_{\alpha } -
\epsilon_i|$, eigenfunctions with larger $|E_{\alpha }|$ 
have larger perturbative parts and so have more $p_j$ 
satisfying Eqs. (\ref{cin2}) or (\ref{cde2}).

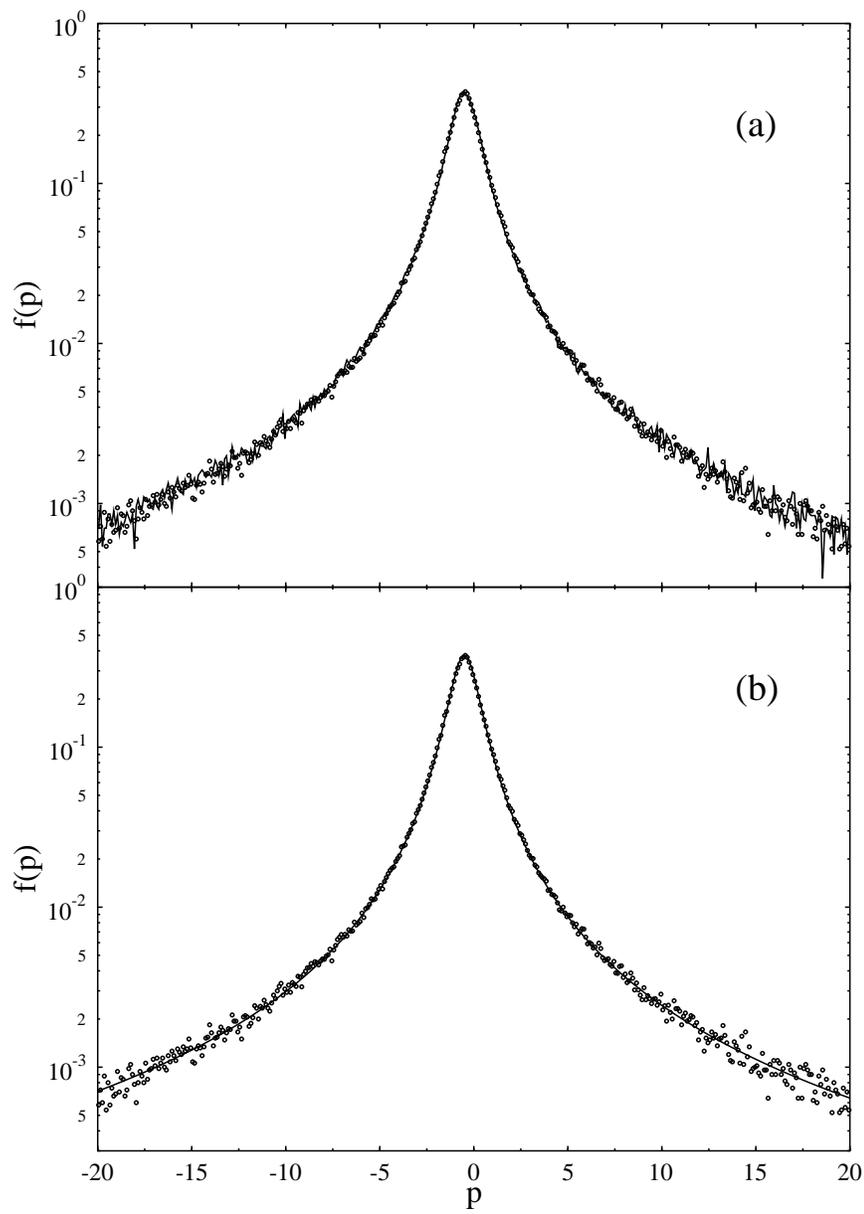
\begin{figure} \caption{(a) A comparison 
(in logarithm scale) between the distribution 
$f_+(p)$ for an increasing eigenfunction 
with $E \approx -4.0044$ (circles) and the $f_-(p)$ for a 
decreasing eigenfunction with $E \approx -4.002$ (solid 
line) when $v=4.0$ and $N = 5 \times 10^{5}$ for the 
Anderson model. (b) $f_+(p)$ in (a) and the fitting 
curve of the Lorentzian form
} \end{figure}
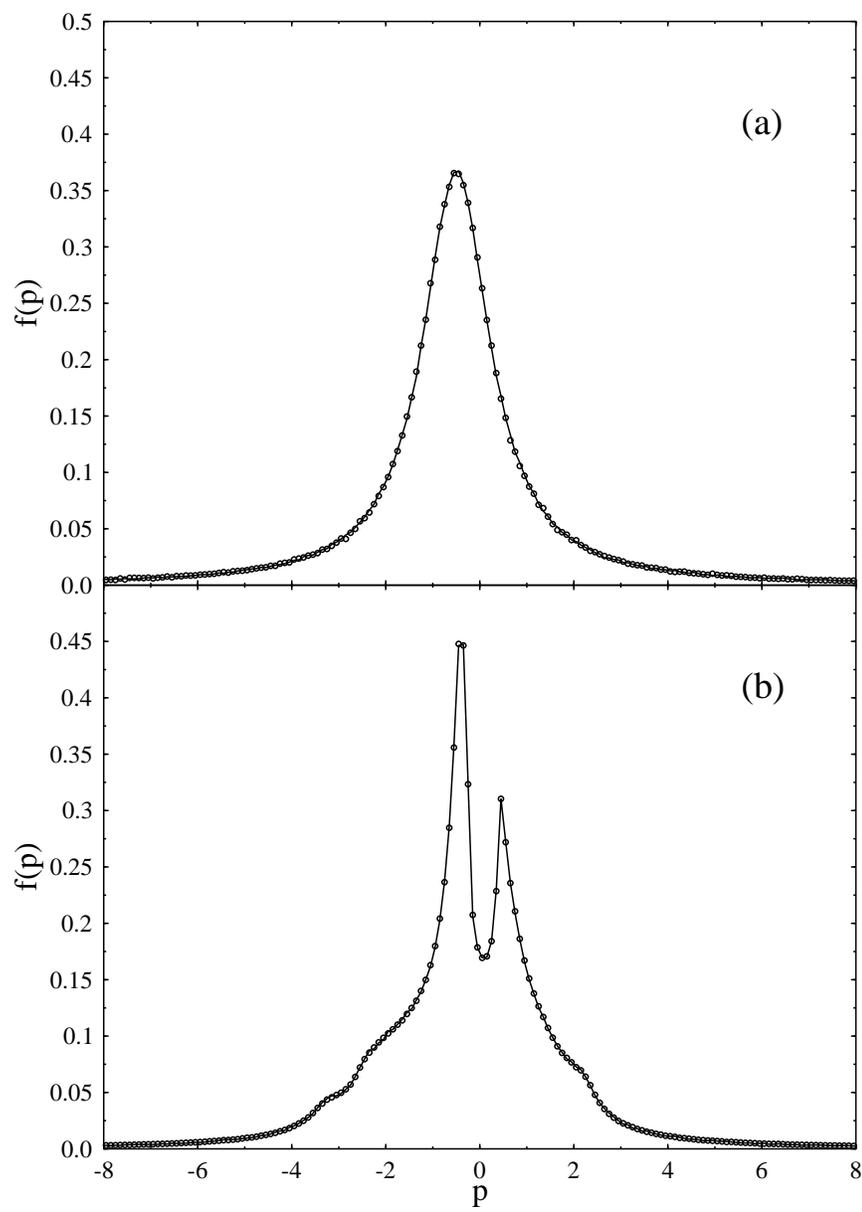
\begin{figure} \caption{ Comparisons between 
$f_a(p)$ obtained by the second method of 
diagonalizing many Hamiltonian matrices (circles) 
and that by the third method of making use of Eq. (3) 
(solid lines) for (a) $v=1.0, E=-1.0$ and (b) 
$v=0.2, E=-0.1$. }
\end{figure} 
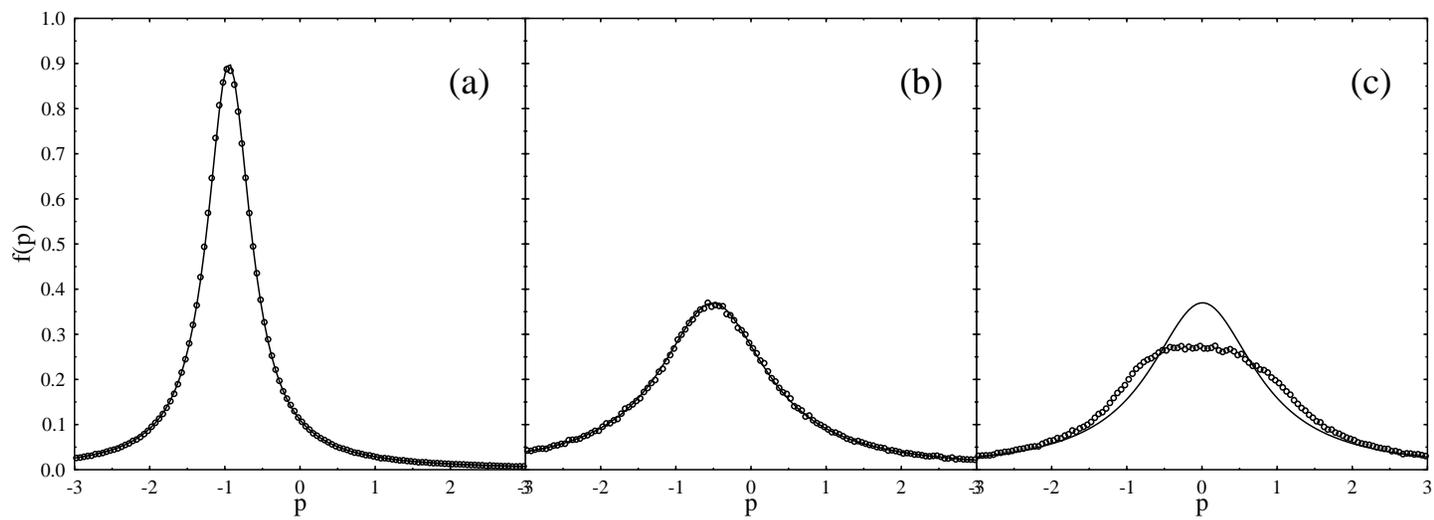
\begin{figure} \caption{Central parts of $f_+(p)$ (circles)
for the Anderson model
and the fitting curves (solid lines)
of the Lorentzian form for $v=4.0$ 
and (a) 
$E=-7.5$, (b) $E=-4.0$, (c) $E=0$. }
\end{figure} 
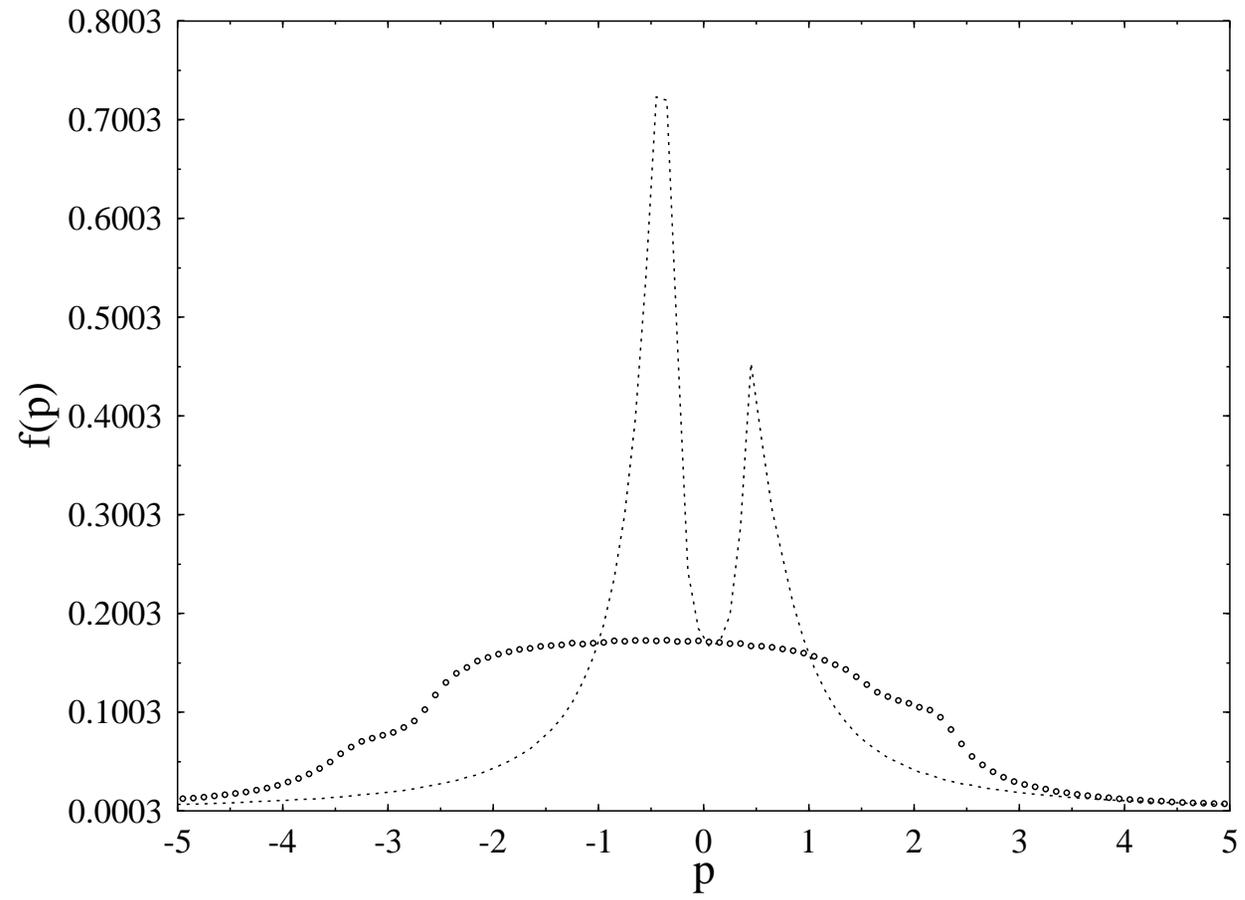
\begin{figure} \caption{Central parts of $f_+(p)$ 
(circles) and $f_-(p)$ (dashed line) for $v=0.2$ and 
$E = -0.1$ for the Anderson model. } 
\end{figure} 
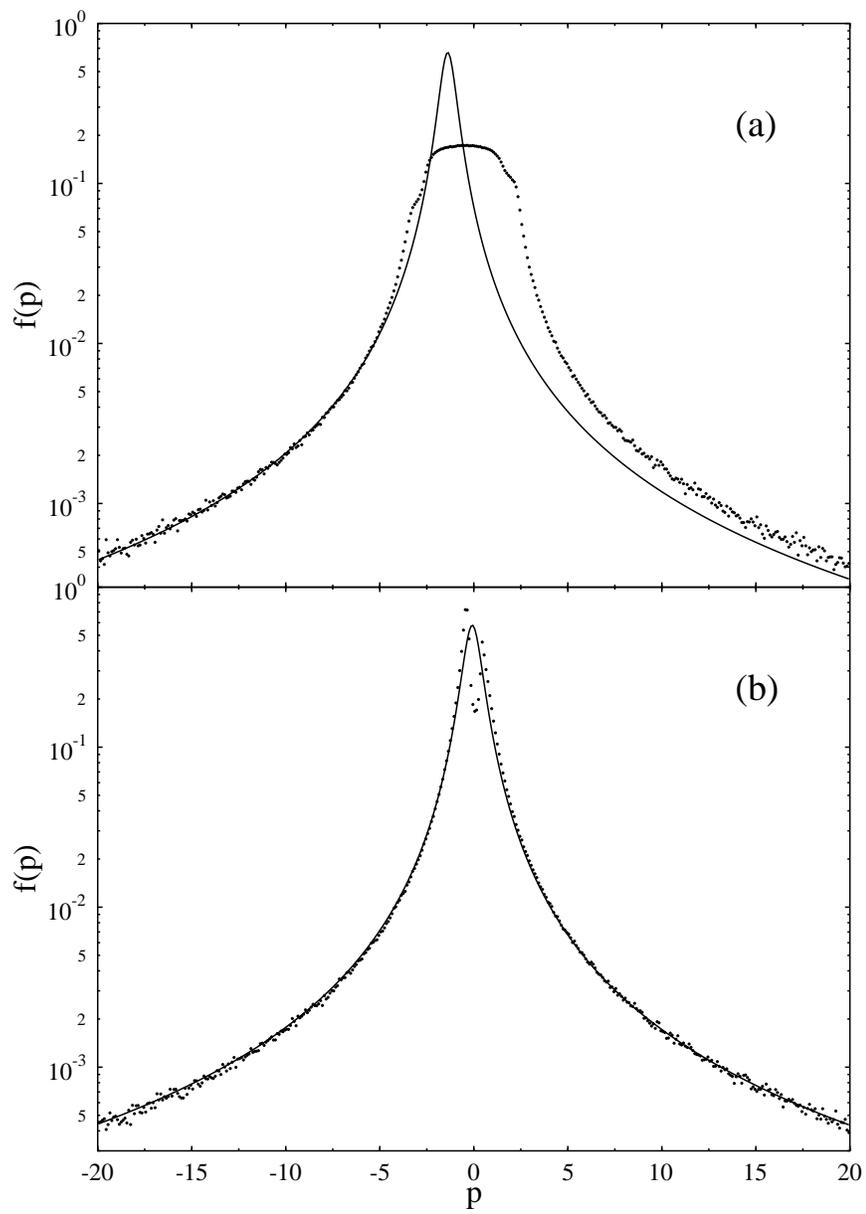
\begin{figure} \caption{ (a) The fitting 
curve (solid line) of the Lorentzian form for the left 
tail (dots) of the $f_+(p)$ in Fig.4 
(in logarithm scale). (b) Same as (a) for tails of 
the $f_-(p)$ in Fig.4. }
\end{figure} 
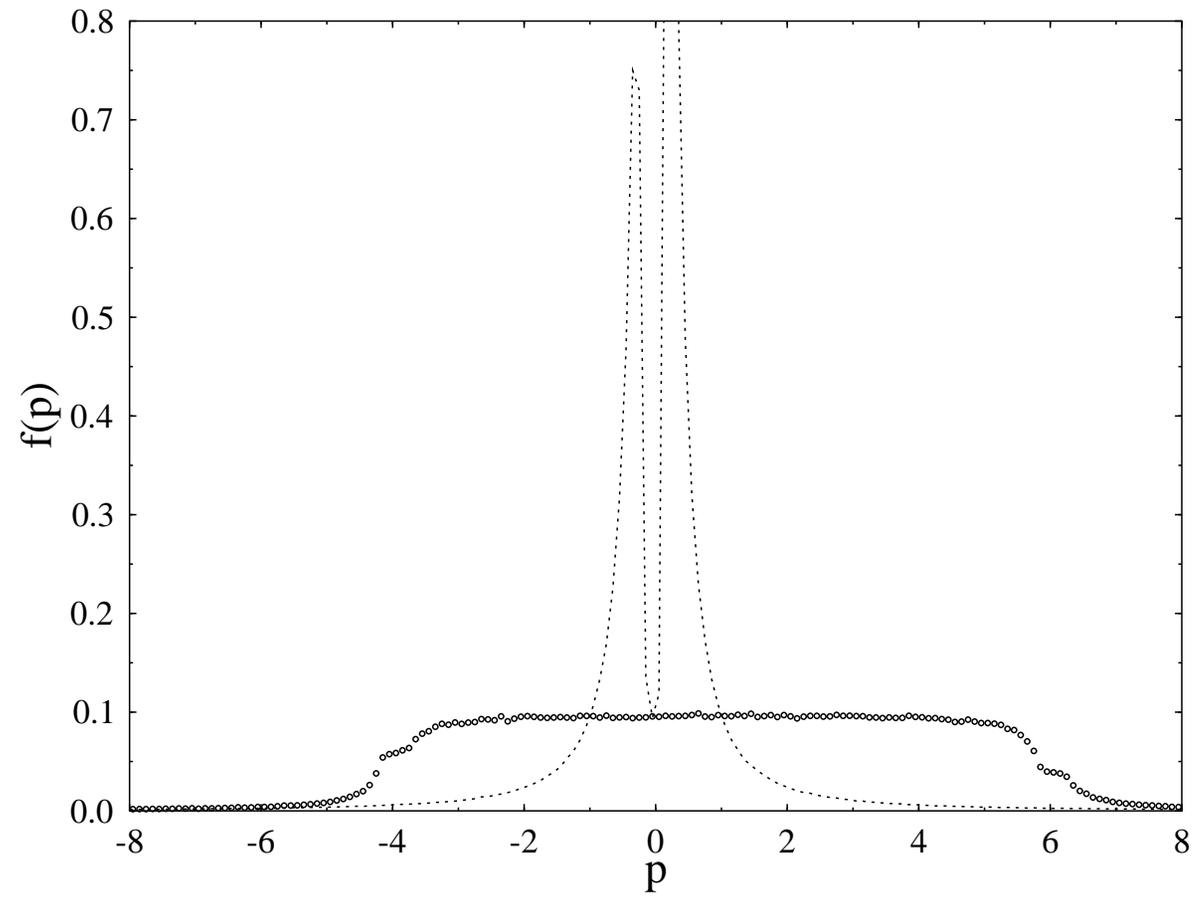
\begin{figure} \caption{Central parts of $f_+(p)$ (circles)
 and $f_-(p)$ (dashed line) for $v=0.1$ and $E=0.1$ for 
the Anderson model. }
\end{figure}
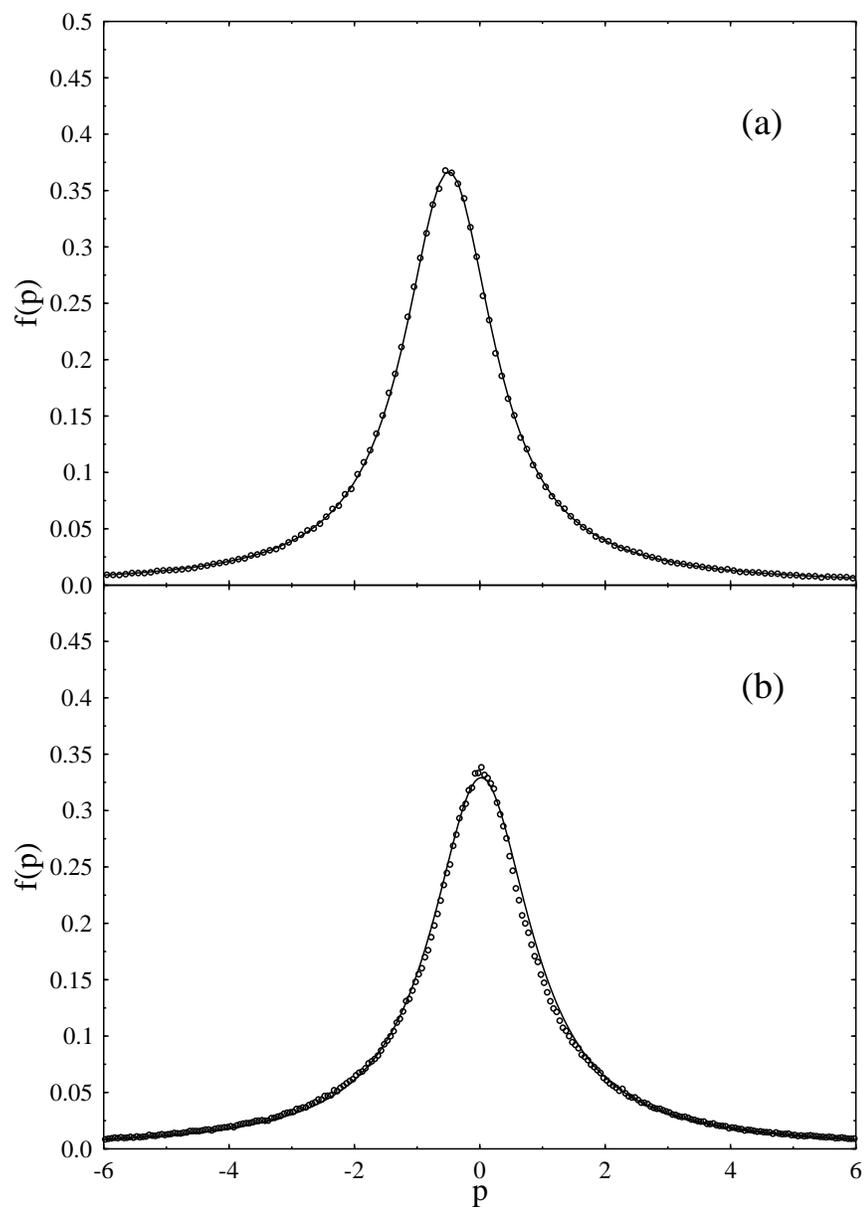
\begin{figure} \caption{ 
Central parts of $f_+(p)$ (circles) 
for the paired correlated Anderson model
and the fitting curves (solid lines) of the 
Lorentzian form for $v=4.0$ and 
 (a) $E=-4.0$, (b) $E=0$. }
\end{figure} 
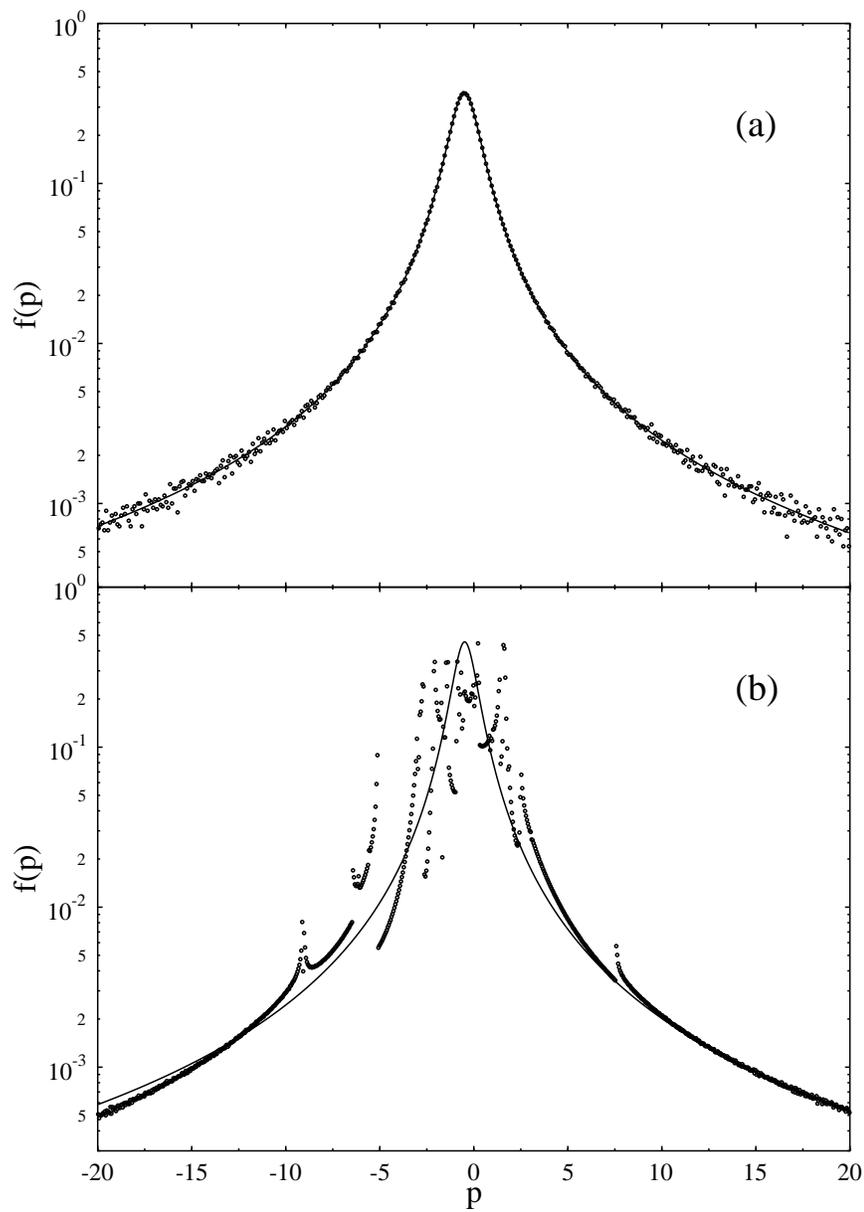
\begin{figure} \caption{ $f_+(p)$ (circles) 
for the  1D quasi-periodic tight binding model 
and the fitting 
curves (solid lines) of the Lorentzian form. 
(a) $v=4.0, E=-4.0$, (b) $v=0.2,E=-0.1$ and the fitting 
curve is for the far right tail only. } 
\end{figure}

\end{document}